# The Structural Phase Transition in Fe$_{1+\delta}$Se

T. M. McQueen[1], A. J. Williams[1], P. W. Stephens[2], J. Tao[3], Y. Zhu[3], V. Ksenofontov[4], F. Casper[4], C. Felser[4], and R. J. Cava[1]

[1]Department of Chemistry, Princeton University Princeton NJ 08544
[2]Department of Physics & Astronomy, SUNY, Stony Brook, New York 11794, USA
[3]Condensed Matter Physics & Materials Science Department, Brookhaven National Laboratory, Upton, NY 11973
[4]Institut für Anorganische Chemie und Analytische Chemie, Johannes Gutenberg-Universität, Staudinger Weg 9, D-55099 Mainz, Germany

In this letter we show that superconducting Fe$_{1.01}$Se undergoes a structural transition at 90 K from a tetragonal to an orthorhombic phase but that non-superconducting Fe$_{1.03}$Se does not. Further, high resolution electron microscopy study at low temperatures reveals an unexpected additional modulation of the crystal structure of the superconducting phase involving displacements of the Fe atoms, and that the non-superconducting material shows a distinct, complex nanometer-scale structural modulation. Finally, we show that magnetism is not the driving force for the phase transition in the superconducting phase.

The discovery of superconducting transition temperatures as high as 55 K in the iron arsenide-based compounds[1-6] has raised numerous questions regarding the underlying physics. The compounds contain structures with layers of edge-sharing FeAs tetrahedra separated by metal ions[1-3, 7, 8]. The undoped compounds, which are non-superconducting, exhibit a tetragonal to orthorhombic structural phase transition[3, 6, 9, 10]. Long range magnetic order, with a moment much reduced from the free Fe$^{2+}$ value, sets in at or slightly below the temperature of that structural transition[3, 9, 10]. On doping[3, 10, 11], the magnetic order and structural transition are suppressed and superconductivity appears. However, the relationship between the structure, magnetism, and superconductivity remains unresolved. For the isomorphic series of compounds LnO$_{1-x}$F$_x$FeAs, reports claim that suppression of both the structural transition and magnetic order (Ln = La)[12], or only the magnetic order (Ln = Ce)[11], or neither (Ln = Sm)[13] is necessary for superconductivity to appear.

The comparatively simple binary compound, tetragonal iron selenide (the "β" form, referred to simply as "FeSe" in the following), has the same basic structure (Fig. 1(a)) and was recently reported to be superconducting at 8.5 K[14]. This compound provides a unique opportunity to study the interplay of the structure, magnetism, and superconductivity in this structure type because of the comparative chemical simplicity: iron selenide has Fe$_2$Se$_2$ layers that are isomorphic to Fe$_2$As$_2$ planes, but lacks intermediate chemical substituents that may affect the electronic and structural properties within the iron layers. Here we report the low temperature structural properties of Fe$_{1.01}$Se (T$_c$ ~ 8.5 K) and Fe$_{1.03}$Se (no T$_c$ > 0.5 K) studied by high resolution synchrotron x-ray powder diffraction (SXRD), transmission electron microscopy (TEM), and electron diffraction (ED). These data show that the structural transition is more complex than previously believed, and that the structural distortion is intimately correlated to the superconductivity. Combined with Mössbauer measurements, these results paint a complex picture of the interplay between structure, magnetism and superconductivity in iron selenide.

The samples were the same as those described previously[15]. SXRD data were collected on the SUNY X16C beamline at the National Synchrotron Light Source. Refinements of the SXRD data were performed using GSAS[16] with the EXPGUI[17] interface. A (001) preferred orientation correction, commonly needed for layered structures, was applied using the March-Dollase method. TEM and ED were performed at room temperature and 11 K on powder samples sitting on copper grids coated with holy carbon in a JEOL 2100F transmission electron microscope equipped with a Gatan liquid helium cooling stage. $^{57}$Fe Mössbauer spectra were recorded in a transmission geometry using a conventional constant-acceleration spectrometer and a helium bath cryostat. The Recoil Mössbauer Analysis Software was used to fit the experimental spectra. Isomer shift values are quoted relative to α-Fe at 293 $K$.

At room temperature, both Fe$_{1.01}$Se and Fe$_{1.03}$Se are well described by the ideal tetragonal unit cell. Rietveld refinements of the room temperature SXRD data of materials of both stoichiometries were carried out, using models containing iron interstitials or selenium vacancies as previously described[15]. The refined formulas were within 2σ of the nominal compositions, irrespective of the model used. That the phases are nearly stoichiometric is consistent with prior neutron diffraction results on the same samples[15]. Since there was negligible difference in the quality of the fits between models with iron interstitials versus selenium deficiencies, for the final refinements, selenium deficiency was assumed and the selenium occupancy was set at the nominal value in each case (i.e. structural formulas of FeSe$_{0.99}$ and FeSe$_{0.97}$ for Fe$_{1.01}$Se and Fe$_{1.03}$Se respectively). The final crystal structure parameters are presented in Table I. They are very similar, with only slight differences in the a- and c- axes, and unit cell volume.

In contrast, Fe$_{1.01}$Se and Fe$_{1.03}$Se display markedly different behavior at low temperature. At 20 K, Fe$_{1.01}$Se

| | $Fe_{1.01}Se$ | | $Fe_{1.03}Se$ | |
|---|---|---|---|---|
| Space group | P 4/n m m | C m m a | P 4/n m m | P 4/n m m |
| Temp. (K) | 298 | 20 | 298 | 20 |
| a (Å) | 3.7727(1) | 5.3100(2) | 3.7787(1) | 3.7682(1) |
| b (Å) | | 5.3344(2) | | |
| c (Å) | 5.5260(3) | 5.4892(2) | 5.5208(2) | 5.4846(2) |
| Vol. (Å³) | 78.652(7) | 155.49(1) | 78.827(6) | 77.877(6) |
| Fe site | 2a (¾,¼,0) | 4a (¼,0,0) | 2a (¾,¼,0) | 2a (¾,¼,0) |
| $B_{iso}$ (Å²) | 1.4(1) | 0.54(6) | 1.3(1) | 0.53(7) |
| Se site | 2c (¼,¼,z) | 4g (0,¼,z) | 2c (¼,¼,z) | 2c (¼,¼,z) |
| z | 0.2668(4) | 0.2665(3) | 0.2666(4) | 0.2673(4) |
| fraction | 0.99 | 0.99 | 0.97 | 0.97 |
| $B_{iso}$ (Å²) | 1.6(1) | 0.68(5) | 1.3(1) | 0.60(5) |
| $\chi^2$ | 2.718 | 1.326 | 2.309 | 2.625 |
| $R_{wp}$ (%) | 16.66 | 15.61 | 13.14 | 15.58 |
| $R_p$ (%) | 12.78 | 12.37 | 10.26 | 11.85 |

**Table 1**. Crystallographic parameters from Rietveld analysis of SXRD data on $Fe_{1.01}Se$ and $Fe_{1.03}Se$. Minor fractions of Fe and/or FeSe(hex) (<3%) were included in the final refinements.

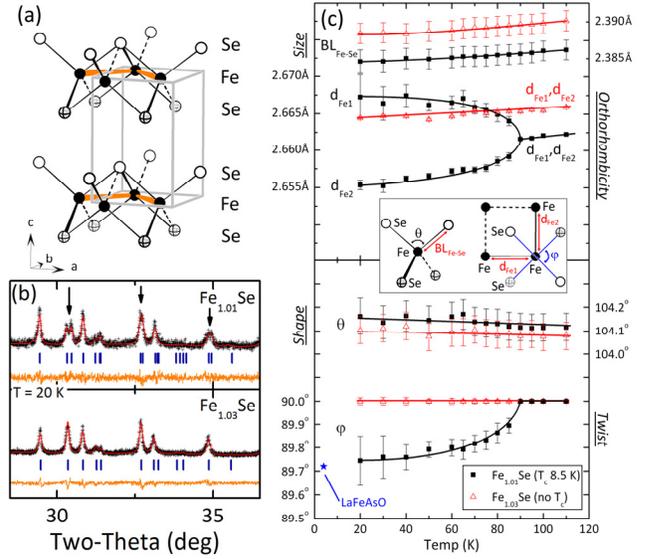

**Fig. 1** (a) The structure of tetragonal iron selenide consists of two-dimensional layers of edge-sharing Fe-Se tetrahedra. (b) Superconducting $Fe_{1.01}Se$ has an orthorhombic distortion, indicated by the splitting of some peaks in SXRD (indicated by arrows), but non-superconducting $Fe_{1.03}Se$ does not. (c) On cooling, $Fe_{1.01}Se$ undergoes a twisting of the selenium tetrahedra, splitting the Fe-Fe distances into two distinct sets. Non-superconducting $Fe_{1.03}Se$, which contains a greater non-stoichiometry and number of defects, shows no transition by SXRD and $d_{Fe1}$ stays equal to $d_{Fe2}$. The torsional angle in the most distorted $Fe_{1.01}Se$ is very similar to that found in undoped (and non-superconducting) LaFeAsO (marked with star).

possesses a lower symmetry structure, evidenced by the splitting of numerous diffraction peaks, whereas $Fe_{1.03}Se$ remains rigorously tetragonal with no peak splitting within the resolution limit of SXRD (Fig. 1(b) and Fig. 2(a)). The low temperature phase of $Fe_{1.01}Se$ is orthorhombic, space group Cmma, with a $\sqrt{2}x\sqrt{2}$ supercell enlargement in the Basal plane, consistent with recent reports[18, 19]. There is no evidence in SXRD for the triclinic structure that was suggested previously[20]. The orthorhombic structure of $Fe_{1.01}Se$ is analogous to that observed in the parent compounds of the FeAs-based superconductors[9, 10]. The structural distortion leading to orthorhombicity is a coherent twisting (away from the ideal 90°) of the upper and lower Se pairs that make up each Fe-Se tetrahedron, and can be described by five parameters: the torsional angle between the Se pairs ($\varphi$), two Fe-Fe distances ($d_{Fe1}$ and $d_{Fe2}$), the Fe-Se bond length ($BL_{Fe-Se}$), and the upper Se-Fe-Se angle ($\theta$). The temperature-dependence of these parameters is shown in Fig. 1(c). The phase transition in $Fe_{1.01}Se$ occurs near 90 K, consistent with resistivity and thermopower measurements[15]. The Fe-Se bond lengths and Se-Fe-Se bond angles are, within error, the same in $Fe_{1.01}Se$ and $Fe_{1.03}Se$, and there is no significant change in these structural characteristics at the phase transition. In contrast, the torsional angle $\varphi$ in $Fe_{1.03}Se$ is 90° independent of temperature, whereas in $Fe_{1.01}Se$ it changes from 90° at high temperatures to 89.7° at 20 K. This 0.3° change is similar in magnitude to the distortion observed in LaFeAsO (star, Fig. 1(c))[9]. The Fe-Fe distances above 90 K are similar in both compounds. Below the phase transition, one Fe-Fe length in $Fe_{1.01}Se$ ($d_{Fe2}$) shortens considerably and the second length ($d_{Fe1}$) elongates. The result is an average difference in long-short Fe-Fe separation of ~0.012 Å in low temperature $Fe_{1.01}Se$. This is a very small difference, but again is similar in magnitude to the difference in Fe-Fe distances in undoped LaFeAsO (2.855 Å - 2.841 Å = 0.014 Å)[9], even though the absolute Fe-Fe distances are substantially shorter in FeSe (2.66 Å in FeSe vs. 2.83 Å in LaFeAsO).

The SXRD data shows that $Fe_{1.01}Se$ undergoes a transition to an orthorhombic phase at 90 K. The same distortion is observed in undoped and lightly doped FeAs-based compounds, and attributed as arising from the magnetic ordering that sets in at or just below the transition[10]. In FeSe, however, no magnetic ordering is observed: Mössbauer spectra (Fig. 2(b)) show no peak splitting or other significant changes through the phase transition, as would be expected if magnetic order was present. At most, from the perspective of magnetism, the present data say that magnetic fluctuations in $Fe_{1.01}Se$, if present, must be on a timescale faster than that of the Mössbauer effect ($10^{-7}$ s). This implies that the structural phase transition in the Fe-based systems is not magnetically driven. Similarly, certain FeAs-systems, at intermediate doping, show a structural distortion but no magnetic order[11]. In others, the structural and magnetic ordering occur simultaneously[21], or are separated in temperature[9]. This decoupling of the magnetic and structural behavior implies that the crystallographic phase transition and magnetic ordering are driven by different effects.

Electron diffraction (ED) patterns at low temperature show that the structural transition is more complex in superconducting $Fe_{1.01}Se$ than expected. At room temperature, the ED patterns are consistent with the ideal tetragonal structures found by SXRD (Fig. 3(a)). However, at low temperature, additional superreflections appear (Fig. 3(b)). The presence of these reflections, which

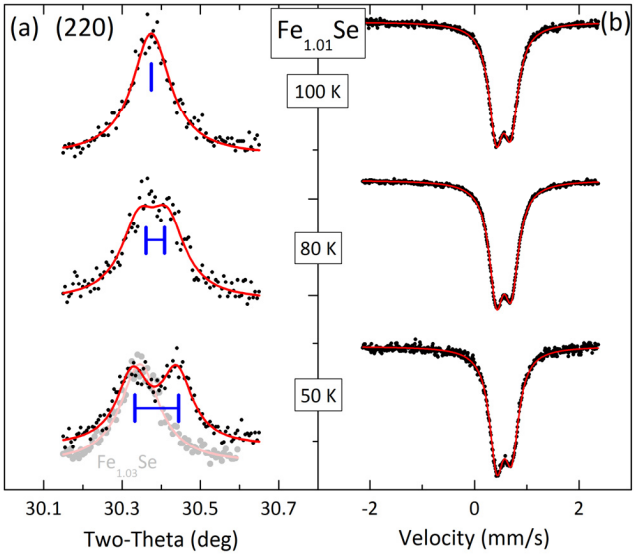

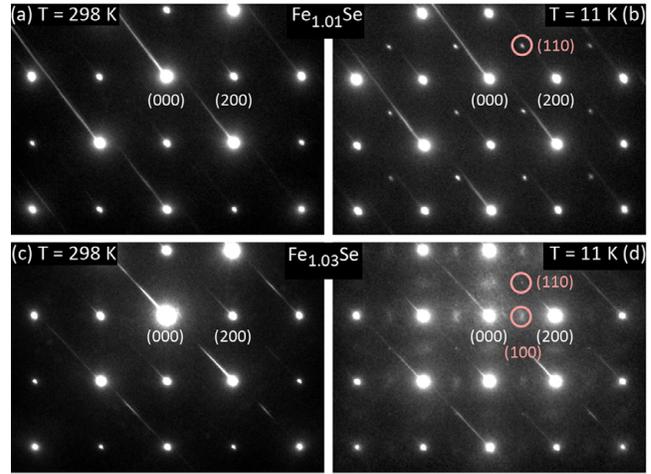

**Fig. 2** (a) SXRD scans of the (220) reflection of $Fe_{1.01}Se$ shows the appearance of the orthorhombic structural distortion near 90 K. The SXRD pattern at 50 K of $Fe_{1.03}Se$, which does not show the distortion, is also shown. (b) Mössbauer spectra of $Fe_{1.01}Se$ are unchanged as the temperature is lowered through the structural distortion, eliminating the onset of long range magnetic order as a possible origin of the transition.

**Fig. 3** (a-d) Electron diffraction patterns of $Fe_{1.01}Se$ and $Fe_{1.03}Se$, indexed with the orthorhombic cell. The room temperature patterns ((a) and (c)) are consistent with the ideal tetragonal cell. (b) Weak superreflections are visible in $Fe_{1.01}Se$ at the (hk0), h+k=2n, h,k odd positions at T = 11 K, indicating a subtle deviation from the orthorhombic structure found by SXRD. (d) $Fe_{1.03}Se$ also shows scattering at those positions, and there is also scattering intensity at the (h00), h = odd and (0k0), k = odd, positions. This scattering is systematically absent in $Fe_{1.01}Se$, implying a more complex modulation in $Fe_{1.03}Se$. The 45° streaks are due to the shutter during the short exposure time used.

appear at all (hk0), h+k = 2n, h,k = odd (e.g. (110)), is surprising. They are not consistent with the Cmma symmetry found by SXRD, which requires that (hk0), h,k = 2n. Multiple scattering, which could explain this discrepancy, cannot be the origin of the extra reflections, as the scattering is only present below the phase transition and both patterns were taken from the same sample area. Instead, the presence of these reflections indicates that the actual low temperature structure of superconducting $Fe_{1.01}Se$ has a subtle departure from Cmma symmetry.

Unexpected extra reflections are also observed in ED for non-superconducting $Fe_{1.03}Se$ at T = 11 K (Fig. 3(d)). The extra reflections are not indexable using the tetragonal unit cell found by SXRD. A $\sqrt{2}x\sqrt{2}$ supercell enlargement in the Basal plane (like in orthorhombic $Fe_{1.01}Se$) is needed. In this expanded cell, the extra reflections occur not only at all (hk0), h+k=2n, h,k = odd positions, as in $Fe_{1.01}Se$, but also at (h00), h = odd and (0k0), k = odd. This is despite no observable lowering of symmetry by SXRD.

Real space images obtained by TEM at low temperatures, shown in Fig. 4(a,b), were used to further investigate the subtle structural modulations. For $Fe_{1.01}Se$, there are closely spaced lattice fringes that are highly aligned and ordered over large areas (more than 50 nm). $Fe_{1.03}Se$, however, does not show such long range uniformity. Some regions appear to be tetragonal, with bidirectional fringes with the same spacing as in $Fe_{1.01}Se$. Other areas have striped fringes along one direction, like in $Fe_{1.01}Se$, but with approximately twice the spacing. These regions are small (c.a. 5 nm), and form a checkerboard-type structural modulation. Fast Fourier Transforms of different regions of a TEM micrograph of $Fe_{1.03}Se$ show that both sets of extra reflections (compare cf Fig. 3(d)) occur simultaneously and come from regions of the sample with the double-sized fringes. This implies that the ordering that gives rise to the superreflections in $Fe_{1.03}Se$ occurs within the nanosized domains. The nanometer size of the ordered structural domains in $Fe_{1.03}Se$ is consistent with the disruption of long range ordering due to the structural defects that must be present in material of this stoichiometry. However, successive warming and cooling of the sample shows that the nanodomains form in different places on each cooling cycle, implying that they are not pinned to defects. This means that the defects in $Fe_{1.03}Se$ are doing more than breaking up the long range order of the structural transition.

The present data does not allow for unambiguous assignment of the origin of the lowering of symmetry in $Fe_{1.01}Se$ or the exact nature of the nanometer-scale structural distortion in $Fe_{1.03}Se$. Some general conclusions can be drawn, however, from crystal-chemical reasoning. For both $Fe_{1.01}Se$ and $Fe_{1.03}Se$, two sets of in-plane reflections (indexed according to $Fe_{1.01}Se$'s orthorhombic supercell in both cases) should be systematically absent: (hk0) h+k=2n, h,k odd, and [(h00), h=odd and (0k0), k=odd]. The first of these conditions comes from the presence of a glide plane that runs through the iron atoms within a layer. The second condition reflects the presence of C-centering, or the translational symmetry of iron atoms within the supercell (Fig. 5(a)). The low temperature ED of $Fe_{1.01}Se$ shows that only the first of these two reflection conditions is violated. This implies that the true symmetry of $Fe_{1.01}Se$ lacks the glide plane but still has the C-centering. The magnitude of the distortion causing this lowering of symmetry must be subtle, as the intensity of the superreflections is <<1% of the primary reflections in the ED patterns, and they are not observed by SXRD. Fig. 5(b and c) shows two ways in which this can occur. The first is by displacement of pairs of iron ions along the short in-plane a-axis. This is consistent with the formation of Fe-Fe dimers, which would imply that the transition is driven by an increase in metal-metal bonding. The second is by displacement of pairs of iron ions along the long in-plane b-axis. This is consistent with an electrostatic effect to avoid a shortened Fe-Fe distance along the short axis. Both could also be occurring simultaneously, resulting in dimers that are twisted off axis (much like the pairs in, e.g., $VO_2$). Additional arrangements are also possible, but these data unequivocally show that the structural phase transition in $Fe_{1.01}Se$ is more complex than previously thought.

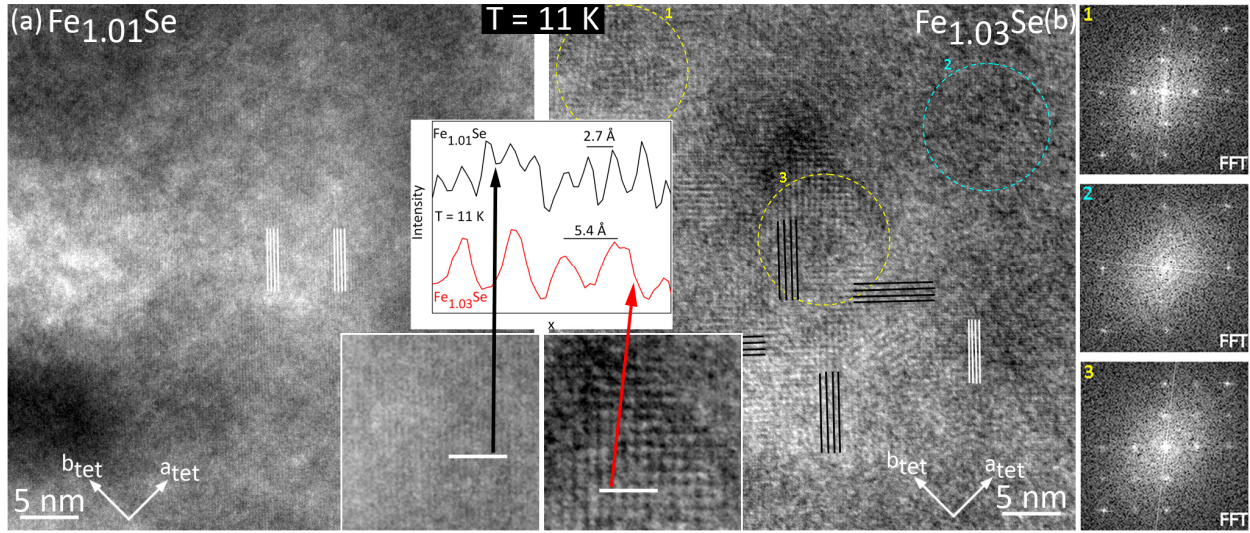

**Fig. 4** TEM micrographs of $Fe_{1.01}Se$ (a) and $Fe_{1.03}Se$ (b) at T = 11 K. $Fe_{1.01}Se$ shows uniform, long range lattice fringes (some marked by vertical lines, also shown in inset). In contrast, $Fe_{1.03}Se$ shows two distinct fringe spacings. The more closely packed type (right-most set of vertical lines) correspond to the undistorted tetragonal structure. The second kind are spaced twice as far apart, and only ordered over short distances (c.a. 5 nm), forming a checkerboard-type structural modulation (some marked by horizontal and vertical lines, also shown in inset). (1,2,3) FFTs of regions of the TEM micrograph of $Fe_{1.03}Se$ show that the scattering at (hk0), h+k=2n, h,k odd, and at (h00), h = odd and (0k0), k = odd arise from the same regions of the sample (1 and 3 show both, 2 shows neither).

A similar complexity is found in $Fe_{1.03}Se$. In this case, both sets of reflection conditions are violated, implying loss of not only the glide plane but also the C-centering. This is consistent with the TEM images (Fig. 4) where fringes are found to be spaced twice as far apart as in $Fe_{1.01}Se$. This cannot simply be due to disordering between adjacent layers stacked along the c axis, as within each layer the C-centering would be preserved. Instead, the loss of C-centering must reflect changes within the plane in addition to those observed in $Fe_{1.01}Se$. One such possibility is shown in Fig. 5(d), where only every other row of iron ions undergoes dimerization. This would break the C-centering, and explain the stripes that are spaced twice as far apart as in $Fe_{1.01}Se$. Regardless of the precise origin, the structural modulation that exists in nanometer size domains in $Fe_{1.03}Se$ is not identical to that found in $Fe_{1.01}Se$. This suggests a link between the observed microstructure and the macroscopic properties: $Fe_{1.01}Se$ superconducts whereas $Fe_{1.03}Se$ does not. Other possibilities to explain what is found in $Fe_{1.03}Se$ include the formation of a charge density wave or $(\pi,\pi)$ electronic order, which would be consistent with related theoretical and experimental results on the iron arsenides[22, 23], but further work is necessary to determine the precise origin.

Low temperature SXRD and Mössbauer data show that superconducting $Fe_{1.01}Se$ undergoes a tetragonal to orthorhombic distortion at 90 K, but without the appearance of long-lived (>$10^{-7}$ s) magnetic order. The distortion itself is analogous to that found in the FeAs-based systems, and is a coherent twisting of the Se pairs that make up the tetrahedra. The presence of the structural transition without magnetic order provides strong evidence that the distortion in these systems is not magnetically driven. The presence of weak superreflections in low temperature ED of superconducting $Fe_{1.01}Se$ indicate a subtle deviation from the structure obtained from SXRD. In contrast $Fe_{1.03}Se$ shows a structural modulation that exists only in nanometer size domains. The nature of the distortions in $Fe_{1.01}Se$ and $Fe_{1.03}Se$ are different, evidence that the excess iron in $Fe_{1.03}Se$ is doing more than simply breaking up the long range coherence of the structural transition. Thus the structural properties of iron-based superconductors, even in this simplest of variants, is more complex than previously envisioned. This work also highlights the importance of local probes in the study of complex phases such as these, as there is no hint of the lower symmetry or domain

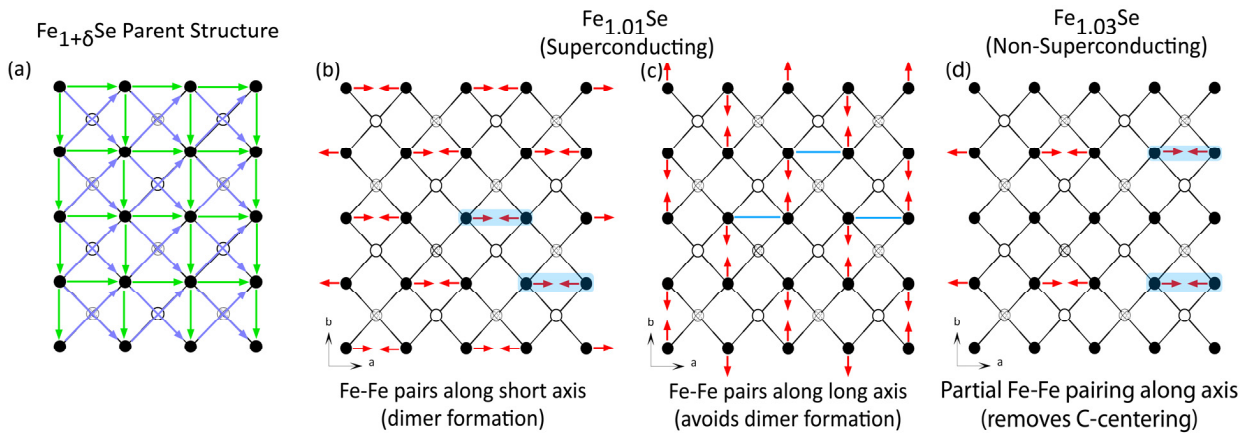

**Fig. 5** (a) The symmetry elements that give rise to in-plane systematic absences in this system are the glide plane (green) and C-centering (light blue), which make different sets of atoms in the unit cell symmetry equivalent. Two ways to break the glide plane but maintain C-centering, as indicated by ED on $Fe_{1.01}Se$, are to displace the iron ions along the in-plane orthorhombic (b) a- or (c) b- axes (or both simultaneously). Displacements along the a-axis are consistent with the formation of dimers (shaded light blue, b), whereas displacements along the b-axis are consistent with the avoidance of a shortened Fe-Fe bond (blue lines, c). $Fe_{1.03}Se$ shows loss of both the glide plane and C-centering, implying that it has an even more complex (but subtle) microstructure. (d) shows one arrangement consistent with the $Fe_{1.03}Se$ data.

structures we observe in ED and TEM for $Fe_{1.01}Se$ and $Fe_{1.03}Se$ when these materials are studied by standard methods such as SXRD or powder neutron diffraction. This suggests that other members of the superconducting iron pnictides should be carefully studied by similar methods, and that until that is done, the subtle relationships between the ubiquitous structural phase transition and superconductivity in this family cannot be resolved.

T.M.M. gratefully acknowledges support by the national science foundation graduate research fellowship program. The work at Princeton was supported by the Department of Energy, Division of Basic Energy Sciences, grant DE-FG02-98ER45706. The work at Brookhaven National Lab (BNL) as well as Use of the National Synchrotron Light Source, BNL, was supported by the U.S. Department of Energy, Office of Science, Office of Basic Energy Sciences, under Contract No. DE-AC02-98CH10886.